\documentclass[reprint, nofootinbib, aps, pra, superscriptaddress]{revtex4-2}

\usepackage{graphicx}
\graphicspath{{./figures/}}
\usepackage[caption=false]{subfig}
\usepackage{hyperref}
\usepackage{amsmath,amsfonts,amssymb,amsthm,bbm,physics,qcircuit}

\usepackage[boxed]{algorithm}
\usepackage{algorithmic}

\newtheorem{Definition}{Definition}
\newtheorem{Theorem}{Theorem}
\newtheorem{Lemma}{Lemma}
\newtheorem{Corollary}{Corollary}

\newcommand{\Cl}{\mathcal{C}\ell}
\newcommand{\Herm}{\text{Herm}}

\begin{document}

\title{Simulation of quantum computation with magic states\texorpdfstring{\\}{}via Jordan-Wigner transformations}

\author{Michael Zurel}
\thanks{These authors contributed equally.}
\affiliation{Department of Physics \& Astronomy, University of British Columbia, Vancouver, Canada}
\affiliation{Stewart Blusson Quantum Matter Institute, University of British Columbia, Vancouver, Canada}
\affiliation{Department of Mathematics, Simon Fraser University, Burnaby, Canada}

\author{Lawrence Z. Cohen}
\thanks{These authors contributed equally.}
\affiliation{Centre for Engineered Quantum Systems, School of Physics, University of Sydney, Sydney, Australia}

\author{Robert Raussendorf}
\affiliation{Stewart Blusson Quantum Matter Institute, University of British Columbia, Vancouver, Canada}
\affiliation{Institute of Theoretical Physics, Leibniz University Hannover, Hannover, Germany}

\date{\today}

\begin{abstract}
    Negativity in certain quasiprobability representations is a necessary condition for a quantum computational advantage. Here we define a new quasiprobability representation exhibiting this property with respect to quantum computations in the magic state model. It is based on generalized Jordan-Wigner transformations, and it has a close connection to the probability representation of universal quantum computation based on the $\Lambda$ polytopes. For each number of qubits, it defines a polytope contained in the $\Lambda$ polytope with some shared vertices. It leads to an efficient classical simulation algorithm for magic state quantum circuits for which the input state is positively represented, and it outperforms previous representations in terms of the states that can be positively represented.
\end{abstract}

\maketitle

\section{Introduction}

Quasiprobability representations serve as a bridge between classical and quantum descriptions of physical systems, with negativity serving as an indicator of genuinely quantum behaviour and the fragment of quantum theory that is positively represented behaving more classically~\cite{Wigner1932,Hudson1974,KenfackZyczkowski2004}.

Gross' discrete Wigner function~\cite{Gross2006,Gross2008}---a quasiprobability representation for systems of odd-dimensional qudits---has been particularly useful in describing quantum computations in the magic state model. In fact, Veitch et al.~\cite{VeitchEmerson2012} showed that negativity in this representation is a necessary condition for a quantum computational advantage over classical computation. This result is obtained by defining an efficient classical simulation algorithm for magic state quantum computations that applies whenever the input state of the quantum circuit has a nonnegative representation. In the last decade, many other quasiprobability representations have been defined which also exhibit this property~\cite{DelfosseRaussendorf2015,PashayanBartlett2015,HowardCampbell2017,KociaLove2017,RallKretschmer2019,RaussendorfZurel2020,SeddonCampbell2021}.

This viewpoint relating negativity and quantum computational advantage was disrupted in Refs.~\cite{ZurelRaussendorf2020,ZurelHeimendahl2021} where a fully probabilistic model describing universal quantum computation was defined. In this model, all quantum states are represented by a probability distribution over a finite set of hidden states, and all computational dynamics are represented by stochastic state update rules---no negative quasiprobabilities are required. However, this circumvention of negativity comes at a cost: it can no longer be guaranteed that the update rules are efficiently computable classically. Thus, although this model yields a classical simulation algorithm for universal quantum computation, the simulation is likely inefficient in general (as would be expected for any classical simulation of universal quantum computation).

\begin{figure}[b]
    \centering
    \vspace{-6mm}
    \includegraphics[width=0.3\textwidth]{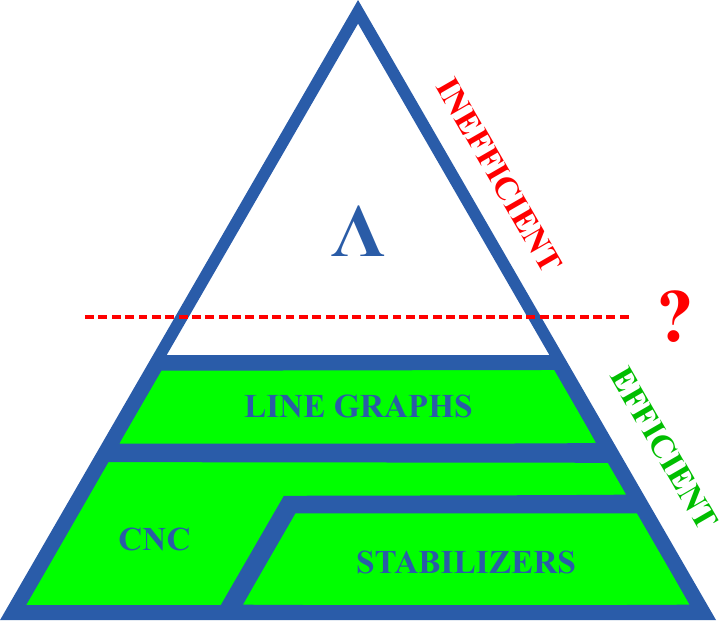}
    \vspace{-3mm}
    \caption{The big question about the $\Lambda$ polytopes~\cite{ZurelRaussendorf2020,ZurelHeimendahl2021} is where the line between efficient and inefficient classical simulation of quantum computation lies. Here we increase the size of the known efficiently simulable region. Our construction is based on Jordan-Wigner transformations, and its range of applicability includes the earlier CNC construction~\cite{RaussendorfZurel2020}, which in turn includes the stabilizer formalism~\cite{Gottesman1998,AaronsonGottesman2004,HowardCampbell2017}.}
    \label{Figure:LambdaHierarchyCartoon}
\end{figure}

The ultimate question for these classical simulation methods based on quasiprobability representations is: \emph{Where runs the line between efficient and inefficient classical simulation, and which physical or mathematical property determines it?} For qubits, so far it is known that this dividing line lies somewhere between the quantum computations covered by the so-called CNC (closed and noncontextual) construction~\cite{RaussendorfZurel2020}, and the universal quantum computations described by the $\Lambda$ polytopes~\cite{ZurelRaussendorf2020} (see Fig.~\ref{Figure:LambdaHierarchyCartoon}).

In this work, we enlarge the known region of efficient classical simulability inside the $\Lambda$-polytopes. Specifically, we define a new quasiprobability representation intermediate between those of the CNC construction and the $\Lambda$-polytopes. This model has efficiently computable state update rules under the dynamics of quantum computation, and it yields an efficient classical simulation algorithm for quantum computations on the subset of input states that are positively represented. It subsumes the CNC construction and previous methods such as those based on quasimixtures of stabilizer states~\cite{HowardCampbell2017}. Thus, we effectively push out the boundary of quantum computations that can be efficiently simulated classically.

The model is based on generalized Jordan-Wigner transformations~\cite{ChapmanFlammia2020}, and its conception was influenced by the surprising connection to Majorana fermions first realized in the CNC construction~\cite{RaussendorfZurel2020}. It was also partially inspired by the techniques of mapping to free fermions in the simulation of other computational models such as matchgate circuits~\cite{JozsaMiyake2008}, which have recently received renewed interest~\cite{MocherlaBrowne2023,CudbyStrelchuk2023,DiasKoenig2023,ReardonSmithKorzekwa2023}.

\medskip

The remainder of this paper is structured as follows. We begin in Section~\ref{Section:Preliminaries} by introducing some notation and definitions. In Section~\ref{Section:PhaseSpaceDescription} we define the new quasiprobability representation of quantum computation with magic states. In Section~\ref{Section:ExtendedClassicalSimulation} we describe the behaviour of the generalized phase space over which this representation is defined with respect to the dynamics of quantum computation with magic states, namely, Clifford gates and Pauli measurements, and we define a classical simulation algorithm for quantum computation with magic states that applies whenever the input state of the quantum circuit is positively represented. We also introduce a monotone for the resource theory of stabilizer quantum computation~\cite{VeitchEmerson2014} for the case where the state is not positively represented. In Section~\ref{Section:NewLambdaVertices} we elucidate the relationship between this model and the $\Lambda$ polytopes. Finally, we conclude with a discussion of these results in Section~\ref{Section:Discussion}.

\section{Preliminaries}\label{Section:Preliminaries}

\begin{figure}
    \centering
    \begin{align*}
    \Qcircuit @C=2em @R=2em {
        \lstick{\ket{H}} & \qw & \ctrl{1} & \qw & \gate{SX} & \qw & \rstick{T\ket{\psi}} \\
        \lstick{\ket{\psi}} & \qw & \targ & \qw & \meter\cwx{-1}
    }
    \end{align*}
    \caption{Implementation of a $T$-gate via injection of the magic state $\ket{H}=\frac{1}{\sqrt{2}}(\ket{0}+e^{i\pi/4}\ket{1})$~\cite[\S10.6]{NielsenChuang2000}. Since the Clifford+$T$ gate set is universal for quantum computation~\cite[\S4.5]{NielsenChuang2000}, this proves universality of the magic state model, which allows Clifford gates and Pauli measurements supplemented with nonstabilizer input states.}
    \label{Figure:Magic state circuit}
\end{figure}
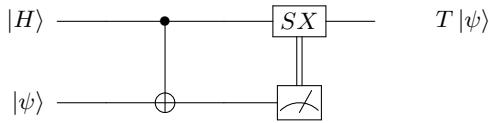

The setting of this paper is quantum computation with magic states (QCM) on systems of qubits~\cite{BravyiKitaev2005}.  In this model, computation proceeds through the application of a sequence of Clifford gates and Pauli measurements on an initially prepared ``magic'' input state. For example, Fig.~\ref{Figure:Magic state circuit} shows the standard implementation of a $T$ gate in this model. In general, the input can be any nonstabilizer quantum state, but for universal quantum computation it suffices to consider input states formed from multiple copies of a single-qubit nonstabilizer state~\cite{Reichardt2009}. The output of the computation is derived from the outcomes of the measurements.

\subsection{Notation}

Before proceeding we need to introduce some notation. The single-qubit Pauli group is the group generated by the Pauli operators $\langle X,Y,Z\rangle$. More generally, the $n$-qubit Pauli group $\mathcal{P}$ is the group of Pauli operators acting on $n$ qubits. It is constructed from tensor products of single-qubit Pauli operators. Quotienting out overall phases we have $\mathcal{P}/\mathcal{Z}(\mathcal{P})\cong\mathbb{Z}_2^{2n}$ and we can fix a phase convention for the Pauli operators to be
\begin{equation}\label{Equation:PauliOperator}
    T_a=i^{-\braket{a_z}{a_x}}Z(a_z)X(a_x)
\end{equation}
for each $a=(a_z,a_x)\in\mathbb{Z}_2^n\times\mathbb{Z}_2^n=:E$, where $Z(a_z)=\bigotimes_{k=1}^{n}Z^{a_z[k]}$, $X(a_x)=\bigotimes_{k=1}^{n}X^{a_x[k]}$, and the inner product $\braket{a_z}{a_x}$ is computed modulo $4$.  The symplectic product $[\cdot,\cdot]:E\times E\rightarrow\mathbb{Z}_2$ defined as
\begin{equation}
    [a,b]=\braket{a_z}{b_x}+\braket{a_x}{b_z},\;\forall a,b\in E
\end{equation}
tracks the commutator of the Pauli operators as $T_aT_b=(-1)^{[a,b]}T_bT_a$. We define a function $\beta$ that tracks the signs that get picked up by composing pairs of commuting Pauli operators as
\begin{equation}\label{Equation:BetaDefinition}
    T_aT_b=(-1)^{\beta(a,b)}T_{a+b}.
\end{equation}

A projector onto the eigenspace of a set of pair-wise commuting Pauli observables $I\subset E$ corresponding to eigenvalues $\{(-1)^{r(a)}\;|\;a\in I\}$ is given by
\begin{equation}
	\Pi_I^r:=\frac{1}{|I|}\sum\limits_{a\in I}(-1)^{r(a)}T_a.
\end{equation}
The set $I\subset E$ and the function $r:I\rightarrow\mathbb{Z}_2$ must satisfy a set of conditions in order for $\Pi_I^r$ to be a valid projector. First, $I$ must be a closed subspace of $E$. Furthermore, in order for the observables in $I$ to be simultaneously measurable they must commute, and so $I$ must be an isotropic subspace of $E$---a subspace on which the symplectic form vanishes. We must also have $(-1)^{r(0)}T_0=1$, or equivalently with the phase convention chosen above, $r(0)=0$. Lastly, for each $a,b\in I$, from the relation $(-1)^{r(a)}T_a\cdot(-1)^{r(b)}T_b=(-1)^{r(a+b)}T_{a+b}$ we must have
\begin{equation}
	r(a)+r(b)+r(a+b)\equiv\beta(a,b)\mod2.
\end{equation}
In the case of a single Pauli measurement $a\in E$ yielding measurement outcome $s\in\mathbb{Z}_2$, the corresponding projector is $\Pi_a^s=(1+(-1)^sT_a)/2$.

A stabilizer state on $n$-qubits is an eigenstate of a set of $n$ independent and pair-wise commuting Pauli operators~\cite{Gottesman1997,Gottesman1998}. In the case $I\subset E$ is a maximal isotropic subspace, $\Pi_I^r$ is a projector onto a stabilizer state.

The gates of QCM are drawn from the Clifford group---the normalizer of the Pauli group in the unitary group up to overall phases: $\Cl:=\mathcal{N}(\mathcal{P})/U(1)$. The Clifford group acts on the Pauli group as
\begin{equation}
    gT_ag^\dagger=(-1)^{\Phi_g(a)}T_{S_ga},\forall g\in\Cl,\forall a\in E,
\end{equation}
where for each $g\in\Cl$, $S_g$ is a symplectic map on $E$ (i.e. a linear map that preserves the symplectic product), and $\Phi_g$ is a function defined through this relation to track the signs that get picked up in the Clifford group action.

The functions $\beta$ and $\Phi_g$ above have a cohomological interpretation defined in Ref.~\cite{OkayRaussendorf2017} that relates them to proofs of contextuality. This was further elucidated in Ref.~\cite{RaussendorfFeldmann2021} which relates them to Wigner functions describing quantum computation with magic states.

In the following, we use the notation $\mathcal{O}$ for a general subset of $E$ (labels for a subset of Pauli operators), and $\mathcal{O}^*$ for the set $\mathcal{O}\setminus\{0\}$ of nonidentity Pauli operators in $\mathcal{O}$. $\mathcal{O}^\perp:=\{a\in E\;|\;[a,b]=0\;\forall b\in\mathcal{O}\}$ is the symplectic complement of $\mathcal{O}$. We denote by $\Herm(\mathcal{H})$ the space of Hermitian operators on Hilbert space $\mathcal{H}$, and unless otherwise specified, $\mathcal{H}=(\mathbb{C}^2)^{\otimes n}$ is the $n$-qubit Hilbert space. $\Herm_1(\mathcal{H})$ is the affine subspace of $\Herm(\mathcal{H})$ consisting of operators with unit trace, and $\Herm_1^{\succeq0}(\mathcal{H})$ is the subset of $\Herm_1(\mathcal{H})$ consisting of positive semidefinite operators. $\Herm_1^{\succeq0}(\mathcal{H})$ contains density operators representing physical quantum states.

\subsection{Previous quasiprobability representations}

\subsubsection*{The CNC construction}

Recently, a quasiprobability representation for QCM was defined based on noncontextual sets of Pauli observables~\cite{RaussendorfZurel2020}, this is the so-called CNC construction.  Therein, phase space points are identified with pairs $(\Omega,\gamma)$, where $\Omega\subset E$ is a subset of Pauli operators and $\gamma:\Omega\rightarrow\mathbb{Z}_2$ is a function on $\Omega$ satisfying two conditions:\footnote{``CNC'' is for Closed and NonContextual after these conditions.}
\begin{enumerate}
    \item \emph{Closure under inference.} For all $a,b\in\Omega$,
    \begin{equation}
        [a,b]=0\Longrightarrow a+b\in\Omega.
    \end{equation}
    \item \emph{Noncontextuality.\footnote{Implicit in this condition on $\gamma$ there is a constraint on the set $\Omega$, namely, that such a noncontextual value assignment exists. Because of Mermin square-style proofs of contextuality~\cite{Mermin1993}, this is not guaranteed for general subsets of $E$.}} $\gamma$ is a noncontextual value assignment on $\Omega$. I.e. a function $\gamma:\Omega\rightarrow\mathbb{Z}_2$ such that $\forall a,b\in\Omega$ with $[a,b]=0$, we have
    \begin{equation}\label{Equation:NoncontextualityCondition}
    	\gamma(a)+\gamma(b)+\gamma(a+b)\equiv\beta(a,b)\mod2.
    \end{equation}
\end{enumerate}

Then the phase space point operator associated to phase space point $(\Omega,\gamma)$ is defined as
\begin{equation}\label{Equation:CNCOperators}
    A_\Omega^\gamma=\frac{1}{2^n}\sum\limits_{b\in\Omega}(-1)^{\gamma(b)}T_b.
\end{equation}
The terms phase space and phase space point operators are used in analogy with the phase space and phase space point operators of the discrete Wigner function~\cite{Gross2006,Gross2008}.

These operators span the space of Hermitian operators on Hilbert space $\mathcal{H}=(\mathbb{C}^2)^{\otimes n}$ and so any $n$-qubit quantum state $\rho$ can be expanded in these operators as
\begin{equation}\label{Equation:CNCStateRepresentation}
    \rho=\sum\limits_{(\Omega,\gamma)}W_\rho(\Omega,\gamma)A_\Omega^\gamma.
\end{equation}

It was shown in Ref.~\cite[\S IV]{RaussendorfZurel2020} that the admissible pairs $(\Omega,\gamma)$ could be characterized as follows. Let $I\subset E$ be an isotropic subspace and let $\tilde{\Omega}\subset E\setminus I$ be a subset of Pauli operators such that $\left[a,b\right]=1$ for all $a,b\in\tilde{\Omega}^*$. Then the sets $\Omega$ can be expressed as $\Omega=\bigcup_{a\in\tilde{\Omega}}\langle a,I\rangle$. The signs $\gamma$ can be chosen freely on $\tilde{\Omega}^*$ and on a basis of $I$, and then the signs on the rest of $\Omega$ are determined by Eq.~(\ref{Equation:NoncontextualityCondition}).

In simpler terms, the phase space point operators have the form
\begin{equation}\label{Equation:CNCOperatorAlt}
    A_\Omega^\gamma=g(A_{\tilde{\Omega}}^{\tilde{\gamma}}\otimes\ket{\sigma}\bra{\sigma})g^\dagger
\end{equation}
where $g\in\Cl$ is a Clifford group element, $\ket{\sigma}$ is a stabilizer state, all elements of $\tilde{\Omega}^*$ pairwise anticommute, and the signs $\tilde{\gamma}$ on $\tilde{\Omega}^*$ can be chosen freely. Interestingly, a set of pairwise anticommuting Pauli operators has the same algebraic structure as a set of Majorana fermion operators and so, modulo the stabilizer state tensor factor in Eq.~(\ref{Equation:CNCOperatorAlt}), we can view the operator $A_{\tilde{\Omega}}^{\tilde{\gamma}}$ as linear combinations of Majorana operators.

In the case that a state $\rho$ has a representation in Eq.~(\ref{Equation:CNCStateRepresentation}) with $W(\Omega,\gamma)\ge0,\;\forall (\Omega,\gamma)$, the expansion coefficients $W(\Omega,\gamma)$ define a probability distribution over the phase space. Phase space point operators of the form Eq.~(\ref{Equation:CNCOperators}) map deterministically to other phase space point operators under any Clifford gate, and they map to convex mixtures of these operators under Pauli measurements (see Ref.~\cite[\S V]{RaussendorfZurel2020} for details). Further, these state updates can be computed efficiently on a classical computer. Thus, for QCM computations in which the input state of the circuit has a nonnegative representation, all aspects of the computation are represented probabilistically by the model. If samples can be obtained efficiently from the distribution $W(\Omega,\gamma)$, then the model yields an efficient classical simulation of the circuit~\cite[\S VI]{RaussendorfZurel2020}.

\subsubsection*{The \texorpdfstring{$\Lambda$}{Lambda} polytopes}

In Ref.~\cite{ZurelRaussendorf2020}, a probability representation (or a hidden variable model) of quantum computation with magic states was defined. It has the same structure as a quasiprobability representation, except for the fact that no negativity is needed. This representation can be defined for qudits of any Hilbert space dimension~\cite{ZurelHeimendahl2021}, but we state only the multiqubit version here. The state space of the model is based on the set
\begin{equation}
	\Lambda = \left\{ X\in\Herm_1(\mathcal{H})|\Tr(\ket{\sigma}\bra{\sigma}X)\ge0\;\forall\ket{\sigma}\in\mathcal{S} \right\}
\end{equation}
where $\mathcal{S}$ denotes the set of pure $n$-qubit stabilizer states. For each number $n$ of qubits, $\Lambda$ is a bounded polytope with a finite number of vertices~\cite{ZurelHeimendahl2021}. We denote the vertices of $\Lambda$ by $\{A_\alpha\;|\;\alpha\in\mathcal{V}\}$ where $\mathcal{V}$ is an index set. Then the model is defined by the following theorem.
\begin{Theorem}[Ref.~\cite{ZurelRaussendorf2020}; Theorem 1]\label{Theorem:HVM}
	For any number of qubits $n$,
	\begin{enumerate}
		\item Any $n$-qubit quantum state $\rho\in\Herm_1^{\succeq0}(\mathcal{H})$ can be decomposed as
		\begin{equation}
			\rho=\sum\limits_{\alpha\in\mathcal{V}}p_\rho(\alpha)A_\alpha,
		\end{equation}
		with $p_\rho(\alpha)\ge0$ for all $\alpha\in\mathcal{V}$, and $\sum_\alpha p_\rho(\alpha)=1$.
		\item For any $A_\alpha,\;\alpha\in\mathcal{V}$, and any Clifford gate $g\in\Cl$, $gA_\alpha g^\dagger$ is a vertex of $\Lambda$. This defines an action of the Clifford group on $\mathcal{V}$ as $gA_\alpha g^\dagger=:A_{g\cdot\alpha}$ where $g\cdot\alpha\in\mathcal{V}$.
		\item For any $A_\alpha,\;\alpha\in\mathcal{V}$ and any Pauli projector $\Pi_a^s$ we have
		\begin{equation}
			\Pi_a^sA_\alpha\Pi_a^s=\sum\limits_{\beta\in\mathcal{V}}q_{\alpha,a}(\beta,s)A_\beta.
		\end{equation}
		with $q_{\alpha,a}(a,s)\ge0$ for all $\beta\in\mathcal{V}$ and $s\in\mathbb{Z}_2$, and $\sum_{\beta,s}q_{\alpha,a}(\beta,s)=1$.
	\end{enumerate}
\end{Theorem}

This theorem describes a hidden variable model for QCM in which (i)~states are represented by probability distributions $p_\rho$ over $\mathcal{V}$, and (ii)~Clifford gates and Pauli measurements are represented by stochastic maps $g\cdot\alpha$ and $q_{\alpha,a}$. In this model, no negative values are needed in the representation of states, gates, or measurements---it is a true probability representation, but we can no longer guarantee that the updates under Clifford gates and Pauli measurements are efficiently computable classically. Thus, although this representation does give a classical simulation algorithm for any magic state quantum circuit, the simulation is inefficient in general. Analyzing the efficiency of simulation using the $\Lambda$ polytope requires a characterization of its vertices.

To date, only the $\Lambda$ polytopes on one and two qubits have been fully characterized. In addition, some vertices of $\Lambda$ are known for every qubit number. For example, it is known that the CNC-type phase space point operators associated to maximal CNC sets are vertices of the $\Lambda$ polytopes~\cite{Heimendahl2019,ZurelRaussendorf2020}.

\section{Multiqubit phase space from Jordan-Wigner transformations}\label{Section:PhaseSpaceDescription}

Below we define a new quasiprobability representation intermediate between the CNC construction and the full $\Lambda$ polytope model based on newly characterized vertices of $\Lambda$ for every number of qubits. This new model can positively represent more states than the CNC construction (though not all quantum states as in the case of $\Lambda$), and it maintains the property of efficiently computable state update rules for the dynamics of QCM.

\subsection{Line graphs and Jordan-Wigner transformations}\label{Section:LineGraph}

\begin{figure}
    \centering
    \subfloat[]{
        \centering
        \includegraphics[width=0.2\textwidth]{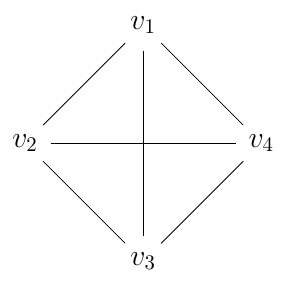}
    }
    \subfloat[]{
        \centering
        \includegraphics[width=0.2\textwidth]{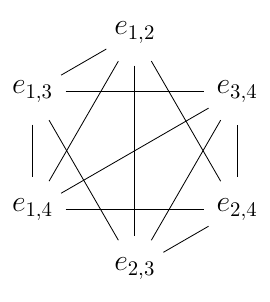}
    }
    \caption{(a) The complete graph on four vertices $K_4$, and (b) the line graph of $K_4$ denoted $L(K_4)$. The edge between vertices $v_i$ and $v_j$ in $K_4$ maps to the vertex $e_{i,j}$ in $L(K_4)$. Two vertices in $L(K_4)$ share an edge if and only if the corresponding edges of $K_4$ share a vertex.}
    \label{Figure:LineGraphExample}
\end{figure}

The generalized phase space we define will consist of operators which can be described by quadratic polynomials of Majorana fermion operators. We motivate this approach by first considering the earlier CNC construction. As alluded to above, on $n$-qubits a maximal CNC set (without stabilizer state tensor factors) is given by $2n+1$ pair-wise anticommuting Pauli operators~\cite[\S IV]{RaussendorfZurel2020}. Such a set of Pauli operators satisfies the anticommutation relations
\begin{equation}
    \{T_a,T_b\}:=T_aT_b+T_bT_a=2\delta_{a,b}
\end{equation}
which are exactly the relations satisfied by a set of Majorana operators. Thus, under a Jordan-Wigner-like transformation, the maximal CNC operators are given by linear combinations of Majorana operators. Furthermore, there is a large body of work studying which operators can be described via quadratic polynomials of Majorana operators. In particular, recently it has been shown that the operators describable in such a way can be identified by the structure of the graphs describing the anticommutation relations of the operators appearing in their Pauli basis expansion~\cite{ChapmanFlammia2020}.

Given a graph $R = (\mathfrak{V}, \mathfrak{E})$, the line graph of $R$, $L(R) = (\mathfrak{E}, \mathfrak{E}')$, is the graph whose vertex set is the edge set of $R$, and whose edge set is $\mathfrak{E}' = \left\{ (e_1, e_2) \in \mathfrak{E} \times \mathfrak{E} | e_1 \cap e_2 \neq \emptyset \right\}$; that is, two vertices in $L(R)$ are neighbours if and only if the corresponding edges in $R$ share a vertex. Given a line graph $L(R)$, we refer to the graph $R$ as the root graph of $L(R)$. See Figure~\ref{Figure:LineGraphExample} for an example. We will also use the notion of twin vertices. Two vertices $u, v \in \mathfrak{V}$ are twin vertices if for every vertex $w \in \mathfrak{V}$, $(u, w) \in \mathfrak{E}$ if and only if $(v, w) \in \mathfrak{E}$.

Let $\mathcal{O} \subset E$ be a subset of Pauli operators. We define the frustration graph of $\mathcal{O}$, $F \left( \mathcal{O} \right) = \left( \mathcal{O}, \mathfrak{E} \right)$, to be the graph with vertices identified with elements of $\mathcal{O}$, and edges drawn between $a, b \in  \mathcal{O}$ if and only if $\left[ a, b \right] = 1$.

\begin{figure}
	\centering
	\subfloat[]{
		\centering
		\includegraphics[width=0.18\textwidth]{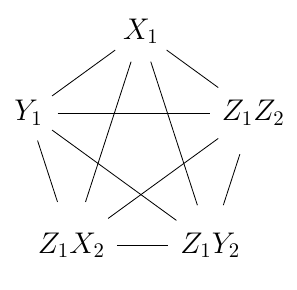}
	}
	\subfloat[]{
		\centering
		\includegraphics[width=0.28\textwidth]{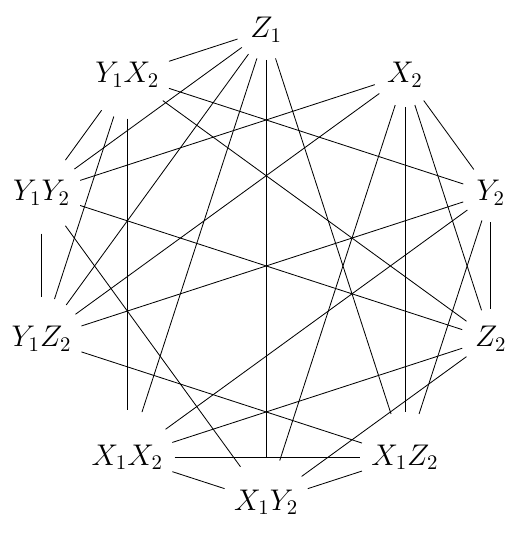}
	}
	\caption{An example of a realization of the line graph $L(K_5)$ as the frustration graph of a set of Pauli operators. (a)~First we find a set of five pair-wise anticommuting Pauli operators, in this case $\{X_1,Y_1,Z_1X_2,Z_1Y_2,Z_1Z_2\}$. The frustration graph of these operators is $K_5$. (b)~The set of Pauli operators obtained by taking all products of pairs of operators in this set has frustration graph $L(K_5)$.}
	\label{Figure:frustration graph specific}
\end{figure}

We now show that every line graph, $L(R)$, can be realized as the frustration graph of some set of Pauli operators $\mathcal{O}$~\cite{ChapmanFlammia2020} (see Fig.~\ref{Figure:frustration graph specific} for an example). To do this, we first construct a set of $|R|$ pair-wise anticommuting Pauli operators. For even $|R|$, we can find such a set by taking the standard Jordan-Wigner transformation of Majorana fermion operators~\cite{JordanWigner1928}
\begin{align}
    C_{2j-1} &= \left(\prod\limits_{k=1}^{j-1} Z_k\right) X_j \label{Equation:JW1} \\
    C_{2j} &= \left(\prod\limits_{k=1}^{j-1} Z_k\right) Y_j \label{Equation:JW2}
\end{align}
for $j = 1, \ldots, |R|/2$. For odd $|R|$, we use Eqs.~\eqref{Equation:JW1} and~\eqref{Equation:JW2} on $(|R|-1)/2$ qubits and also include the operator
\begin{equation}
    C_{|R|} = \prod\limits_{k=1}^{(|R|-1)/2} Z_k.
\end{equation}
Now we identify each vertex $r$ in $R$ with an operator $C_r$. We can then identify each edge $(r,s)$ in $R$, and consequently each vertex in $L(R)$, with a Pauli operator $\pm iC_rC_s$. Two Pauli operators $\pm iC_{r}C_{s}$ and $\pm iC_{t}C_{u}$ anticommute if and only if they share exactly one Majorana operator, but this is equivalent to the associated edges in $R$ sharing a vertex. Thus, the frustration graph of this set of quadratic Majorana operators is exactly the line graph $L(R)$, as desired. Note that this is not necessarily the most efficient way to realize each line graph as a frustration graph, though it does work for any line graph.

\subsection{Definition of the generalized phase space}

In this work, we are interested in operators with the following structure.
\begin{Definition}\label{Definition:LineGraphOperator}
    A Hermitian operator expanded in the Pauli basis as
    \begin{equation}\label{Equation:LineGraphOperator}
        A_{\mathcal{O}}^{c} = \frac{1}{2^n} \left( 1 + \sum_{b \in \mathcal{O}^*} c_b T_b \right)
    \end{equation}
    is a line graph operator if the frustration graph of $\mathcal{O}^*$ is a line graph and $\mathcal{O}^*\cap\mathcal{O}^\perp=\emptyset$.
\end{Definition}

We actually wish to be a bit more permissive than requiring the frustration graphs to be exactly line graphs. In particular, we also want to include operators that have been obtained by tensoring on projectors onto stabilizer states and conjugating by a Clifford operators, as in Eq.~(\Ref{Equation:CNCOperatorAlt}). In this case, the frustration graph will only be a line graph up to twin vertices~\cite{ChapmanFlammia2020}. Suppose we take an operator which is the tensor product of a projector onto an $m$-qubit stabilizer state, $\ket{\sigma}\bra{\sigma}$, and an $(n-m)$-qubit line graph operator, $A_{\mathcal{O}}^{c}$. Then for each nontrivial Pauli operator in the expansion of $A_{\mathcal{O}}^{c}$ there are $2^m$ associated vertices in the frustration graph of $A_{\mathcal{O}}^{c}\otimes\ket{\sigma}\bra{\sigma}$. These vertices will all be pairwise twin vertices. Conversely, the presence of twin vertices alone does not imply that the operator factors through a stabilizer state, there are graphs admitting partitions into equally sized sets of twin vertices that do not correspond to such a tensor product decomposition. In the formalism we include stabilizer states via explicit tensor factors, and these are tracked separately in the classical simulation algorithm described below.

In order for a model constructed from operators with the above structure to correctly reproduce the quantum mechanical predictions for all sequences of Pauli measurements, they must be in the $\Lambda$ polytopes~\cite{ZurelRaussendorf2020}. Otherwise, the model could predict negative quasiprobabilities for some sequences of Pauli measurements, and the model could fail to be closed under Pauli measurement. Without any more constraints, this defines an infinite set of operators for each number of qubits since the coefficients in Eq.~(\ref{Equation:LineGraphOperator}) can vary continuously. For the purpose of defining the generalized phase space of our model, we focus our attention on a finite subset of these operators. We choose this set in the following way. Let $\mathcal{O}$ be a set of Pauli operators such that the frustration graph of $\mathcal{O}^*$ is a line graph and $\mathcal{O}^*\cap\mathcal{O}^\perp=\emptyset$. The operators of the form Eq.~(\ref{Equation:LineGraphOperator}) with support $\mathcal{O}$ form an affine subspace of $\Herm_1(\mathcal{H})$. Intersecting the polytope $\Lambda$ with this affine subspace defines a new polytope with a finite number of vertices, and when interpreted as a polytope embedded in the same space as $\Lambda$ with the Pauli coefficients of $E\setminus\mathcal{O}$ equal to zero, it is contained in $\Lambda$. Let $\mathcal{V}_{\mathcal{O}}$ be the index set for these vertices. We define an initial generalized phase space to be the set $\mathcal{V}_{LG}:=\bigcup_{\mathcal{O}}\mathcal{V}_{\mathcal{O}}$, where we allow the support $\mathcal{O}$ to vary over all inclusion-maximal sets of Pauli operators with line graphs as frustration graphs.

Finally, we define our generalized phase space $\mathcal{V}$ to be the phase space $\mathcal{V}_{LG}$, augmented with stabilizer state tensor factors and conjugation by Clifford operators, analogous to Eq.~\eqref{Equation:CNCOperatorAlt}. This augmentation of the line graph operators is natural for our purposes since tensoring on stabilizer states and conjugating by Clifford operators does not increase the cost of classical simulation of quantum computation~\cite{OkayRaussendorf2017}, and furthermore, arises naturally since we allow Clifford gates and Pauli measurements in the computational model.

To summarize, the set of phase space point operators corresponding to our generalized phase space $\mathcal{V}$ is the set of all operators that can be constructed by the following procedure:
\begin{enumerate}
    \item Start by defining a support $\tilde{\mathcal{O}}$ such that $F(\tilde{\mathcal{O}}^*)$ is a line graph and $\tilde{\mathcal{O}}^*\cap\tilde{\mathcal{O}}^\perp=\emptyset$, and $\tilde{\mathcal{O}}$ is inclusion maximal relative to this property.
    \item Intersect $\Lambda$ with the subspace of $\Herm_1(\mathcal{H})$ spanned by the operators in $\tilde{\mathcal{O}}^*$, choose an extreme point $A_{\tilde{\mathcal{O}}}^{\tilde{c}}$ of this polytope.
    \item Choose a stabilizer state $\ket{\sigma}\in\mathcal{S}$ and a Clifford gate $g\in\Cl$, return 
    \begin{equation}\label{Equation:phasespace}
        A_{\mathcal{O}}^{c}:=g(A_{\tilde{\mathcal{O}}}^{\tilde{c}}\otimes\ket{\sigma}\bra{\sigma})g^\dagger
    \end{equation}
    as the phase space point operator.
\end{enumerate}
Such an operator can be uniquely labeled by its support $\mathcal{O}$ in the Pauli basis and the corresponding Pauli basis coefficients $c$, or equivalently, by the support $\tilde{\mathcal{O}}$ and coefficients $\tilde{c}$ together with a description of the stabilizer state $\ket{\sigma}$ and the Clifford group element $g$.

With these conditions defining the generalized phase space $\mathcal{V}$, any state $\rho$ can be decomposed as
\begin{equation}\label{Equation:LineDecomp}
    \rho = \sum_{(\mathcal{O}, c) \in \mathcal{V}} W_\rho (\mathcal{O}, c) A_{\mathcal{O}}^{c},
\end{equation}
with real coefficients satisfying $\sum_{\mathcal{O},c} W(\mathcal{O},c)=1$. The function $W_\rho(\mathcal{O},c)$ constitutes the representation of states in the model. Note that since the phase point operators are over-complete, this representation is not unique. It is generally preferable to choose a representation that minimizes the amount of negativity as measured by the $1$-norm of the coefficients $W_\rho(\mathcal{O},c)$~\cite{PashayanBartlett2015}. This can be obtained through linear programming.

In the second step of the procedure above defining the generalized phase space $\mathcal{V}$, we take extreme points from an intersection of the relevant $\Lambda$ polytope. Including only extreme points in this way is convenient for the purpose of representing quantum states in Eq.~\eqref{Equation:LineDecomp} since it means $\mathcal{V}$ is finite for each number of qubits. However, this extremal property is not necessary for the classical simulation algorithm in the next section. We introduce the notation $\bar{\mathcal{V}}$ to be the index set corresponding to the set of all operators of the form Eq.~\eqref{Equation:phasespace}, which are in $\Lambda$ and for which the support $\tilde{\mathcal{O}}$ has a line graph structure.

\section{Extended classical simulation}\label{Section:ExtendedClassicalSimulation}

The goal of this section is to show that the generalized phase space constructed in Section~\ref{Section:PhaseSpaceDescription} is closed under the dynamics of quantum computation with magic states---Clifford gates and Pauli measurements. This allows us to define a classical simulation algorithm for quantum computation with magic states in Section~\ref{Section:ClassicalSimulation}.

\subsection{Closure under Clifford gates}\label{Section:CliffordUpdateRules}

First we consider the action of the Clifford group on the generalized phase space $\mathcal{V}_{LG}$. We have the following lemma.

\begin{Lemma}\label{Lemma:CliffordClosureLG}
    For any line graph operator $A_{\mathcal{O}}^{c}$ in $\Lambda$, and any Clifford group element $g\in\Cl$, the operator $gA_{\mathcal{O}}^{c}g^\dagger$ is a line graph operator in $\Lambda$.
\end{Lemma}

\begin{proof}[Proof of Lemma~\ref{Lemma:CliffordClosureLG}]
    Let $A_\mathcal{O}^{c}$ be an operator of the form Eq.~\eqref{Equation:LineGraphOperator}, where the frustration graph of the Pauli operators in $\mathcal{O}^*$ is a line graph and the coefficients $c$ are chosen so that $A_\mathcal{O}^{c}$ is in $\Lambda$.

    For any Clifford operation $g\in\Cl$, we have
    \begin{align}
        gA_\mathcal{O}^{c}g^\dagger=&\frac{1}{2^n}\sum\limits_{b\in\mathcal{O}^*}c_bgT_bg^\dagger\nonumber\\
        =&\frac{1}{2^n}\sum\limits_{b\in\mathcal{O}^*}c_b(-1)^{\Phi_g(b)}T_{S_gb}=:A_{g\cdot\mathcal{O}}^{g\cdot c}.
    \end{align}
    This defines an action of $g$ on the support $\mathcal{O}$ as
    \begin{equation}\label{Equation:CliffordUpdateSupport}
        g\cdot\mathcal{O}=\{S_gb\;|\;b\in\mathcal{O}\},
    \end{equation}
    and on the coefficients $c$ as
    \begin{equation}\label{Equation:CliffordUpdateCoefficients}
        (g\cdot c)_{S_gb}=c_b(-1)^{\Phi_g(b)}.
    \end{equation}
    Since $\Lambda$ is closed under Clifford operations~\cite{ZurelRaussendorf2020}, if $A_\mathcal{O}^{c}$ is in $\Lambda$ then $A_{g\cdot\mathcal{O}}^{g\cdot c}$ is also in $\Lambda$. Since Clifford operations preserve commutation relations of Pauli operators, the frustration graph of $g\cdot\mathcal{O}^*$ is isomorphic to that of $\mathcal{O}^*$. Thus, if the frustration graph of $\mathcal{O}^*$ is a line graph then so is the frustration graph of $g\cdot\mathcal{O}^*$.
\end{proof}

\begin{Corollary}\label{Corollary:CliffordClosure}
    The phase spaces $\mathcal{V}$ and $\bar{\mathcal{V}}$ are each closed under Clifford group operations.
\end{Corollary}

This corollary follows immediately from Lemma~\ref{Lemma:CliffordClosureLG} and from the definitions of $\mathcal{V}$ and $\bar{\mathcal{V}}$. In fact, this corollary follows immediately from the definitions even without Lemma~\ref{Lemma:CliffordClosureLG}. The point of introducing Lemma~\ref{Lemma:CliffordClosureLG} is to note that there is some freedom in how we choose to represent phase space point operators, by absorbing Clifford group elements into the coefficients of the line graph operator part via Eqs.~\eqref{Equation:CliffordUpdateSupport} and~\eqref{Equation:CliffordUpdateCoefficients}. This freedom will become useful in describing the update of the phase space under Pauli measurements in the next section, and in the classical simulation algorithm in Section~\ref{Section:ClassicalSimulation}.

\begin{Corollary}\label{Corollary:CliffordClosureStates}
    The set of states positively representable by $\mathcal{V}$ in Eq.~\eqref{Equation:LineDecomp} is closed under Clifford group operations.
\end{Corollary}

This follows immediately from Corollary~\ref{Corollary:CliffordClosure} by considering the action of a Clifford group element $g\in\Cl$ on a nonnegative decomposition of a state $\rho$ of the form $\rho=\sum_{\alpha\in\mathcal{V}}W_\rho(\alpha)A_\alpha$. Conjugating by $g$ on the left-hand side gives the state $g\rho g^\dagger$, and on the right-hand side we obtain a nonnegative decomposition of $g\rho g^\dagger$ in terms of phase point operators of the phase space $\mathcal{V}$.

\subsection{Closure under Pauli measurements}\label{Section:PauliUpdateRules}

Now we consider the effect of Pauli measurements with respect to the generalized phase space $\mathcal{V}$ (or $\bar{\mathcal{V}}$). First, consider the action of a Pauli projector $\Pi_a^s$, corresponding to a Pauli measurement $T_a,\;a\in E$, yielding measurement outcome $s\in\mathbb{Z}_2$, on a phase space point operator $A_\mathcal{O}^{c},\;(\mathcal{O},c)\in\mathcal{V}_{LG}$ (or $\bar{\mathcal{V}}_{LG}$). This is the content of the following lemma.

\begin{Lemma}\label{Lemma:PauliClosureLG}
    For any line graph operator $A_{\mathcal{O}}^{c}$, $(\mathcal{O},c)\in\mathcal{V}_{LG}$, and any Pauli projector $\Pi_a^s$, we have
    \begin{enumerate}
        \item $\Tr(\Pi_a^sA_{\mathcal{O}}^{c})=\begin{cases}
                (1+(-1)^sc_a)/2,\text{ if $a\in\mathcal{O}$,}\\
                1/2,\text{ otherwise.}
            \end{cases}$
        \item If $\Tr(\Pi_a^sA_{\mathcal{O}}^{c})>0$ then, up to normalization, $\Pi_a^sA_{\mathcal{O}}^{c}\Pi_a^s$ is a phase space point operator of $\bar{\mathcal{V}}$, and if $\Tr(\Pi_a^sA_{\mathcal{O}}^{c})=0$ then $\Pi_a^sA_{\mathcal{O}}^{c}\Pi_a^s=0$
    \end{enumerate}
\end{Lemma}

\begin{proof}[Proof of Lemma~\ref{Lemma:PauliClosureLG}]
    The first statement of the lemma follows from a direct calculation.
    \begin{align}
        \Tr(A_{\mathcal{O}}^{c}\Pi_a^s) =& \frac{1}{2^{n+1}}\sum\limits_{b\in\mathcal{O}}c_b\left[\Tr(T_b)+(-1)^s\Tr(T_aT_b)\right]\nonumber\\
        =& \frac{1}{2}\sum\limits_{b\in\mathcal{O}}c_b\left[\delta_{b,0}+(-1)^s\delta_{b,a}\right]\nonumber\\
        =& \begin{cases}
                (1+(-1)^sc_a)/2,\text{ if $a\in\mathcal{O}$,}\\
                1/2,\text{ otherwise.}
            \end{cases}
    \end{align}
    Here in the second line we use the orthogonality relations of the Pauli operators $\Tr(T_aT_b)=2^n\delta_{a,b}$.

    For the second statement of the lemma, we take each case separately. First, assume $\Tr(\Pi_a^sA_{\mathcal{O}}^{c})>0$. Then conjugating $A_{\mathcal{O}}^{c}$ by the projector $\Pi_a^s$ we have
    \begin{align}
        \Pi_a^sA_{\mathcal{O}}^{c}\Pi_a^s =& \frac{1}{2^{n+2}}\sum\limits_{b\in\mathcal{O}}c_b[T_b+T_aT_bT_a+(-1)^s\left(T_aT_b+T_bT_a\right)]\nonumber\\
        =& \frac{1}{2^{n+1}}\sum\limits_{b\in\mathcal{O}\cap a^\perp}c_b[T_b+(-1)^sT_aT_b]\nonumber\\
        =& \Pi_a^sA_{\mathcal{O}\cap a^\perp}^{c|_{\mathcal{O}\cap a^\perp}}.
    \end{align}
    Using Witt's lemma~\cite{Aschbacher2000}, we can find a Clifford group element $g_1\in\Cl$ (depending on $a$ and $s$) such that $g_1^\dagger\Pi_a^sg_1=\Pi_{z_n}^0$ and $g_1^\dagger A_{\mathcal{O}\cap a^\perp}^{c|_{\mathcal{O}\cap a^\perp}}g_1= A_{\tilde{\mathcal{O}}}^{\tilde{c}} \otimes 1$. Here $A_{\tilde{\mathcal{O}}}^{\tilde{c}}$ is an operator on $n-1$ qubits. Then we have
    \begin{equation}
        \Pi_a^sA_{\mathcal{O}}^{c}\Pi_a^s = \Pi_a^sA_{\mathcal{O}\cap a^\perp}^{c|_{\mathcal{O}\cap a^\perp}} = g_1(A_{\tilde{\mathcal{O}}}^{\tilde{c}} \otimes \Pi_z^0)g_1^\dagger.
    \end{equation}
    In this last expression, $\Pi_z^0$ is a single-qubit Pauli projector. The frustration graph of $\tilde{\mathcal{O}}^*$ is isomorphic to that of $(\mathcal{O}\cap a^\perp)^*$ which is a subgraph of the frustration graph of $\mathcal{O}^*$. Line graphs are closed under induced subgraphs~\cite{Beineke1970}, and so since the frustration graph of $\mathcal{O}^*$ is a line graph, the frustration graph of $\tilde{\mathcal{O}}^*$ is also a line graph. Since $\Lambda$ is closed under Pauli measurements, up to normalization $\Pi_a^sA_{\mathcal{O}}^c\Pi_a^s$ is in $\Lambda$. Thus, $(\tilde{O},\tilde{c})\in\bar{\mathcal{V}_{LG}}$, and the operator $\Pi_a^sA_{\mathcal{O}}^{c}\Pi_a^s$ is a phase space point operator of $\bar{\mathcal{V}}$.

    For the second case, assume that $\Tr(\Pi_a^sA_{\mathcal{O}}^{c})=0$, that is, $c_a=(-1)^{s+1}$. Following the same calculation as in the previous case,
    \begin{equation}\label{Equation:PauliUpdateExpansion}
        \Pi_a^sA_{\mathcal{O}}^{c}\Pi_a^s = \frac{1}{2^{n+1}}\sum\limits_{b\in\mathcal{O}\cap a^\perp}c_b\left[T_b+(-1)^sT_aT_b\right].
    \end{equation}
    I.e., the support of $\Pi_a^sA_{\mathcal{O}}^{c}\Pi_a^s$ is contained in the union of $\mathcal{O}\cap a^\perp$ and $a+(\mathcal{O}\cap a^\perp)$.
    
    For any $d\in\mathcal{O}\cap a^\perp$ and any $t\in\mathbb{Z}_2$, by direct calculation we obtain
    \begin{align}
        &\Tr(\Pi_d^t\Pi_a^sA_{\mathcal{O}}^{c}\Pi_a^s) = \Tr(\Pi_d^t\Pi_a^sA_{\mathcal{O}}^{c}\Pi_a^s\Pi_d^t)\nonumber\\
        &= \frac{1}{2^{n+1}}\sum\limits_{b\in\mathcal{O}\cap a^\perp}c_b\left[\Tr(\Pi_d^tT_b\Pi_d^t)+(-1)^s\Tr(\Pi_d^tT_aT_b\Pi_d^t)\right]\nonumber\\
        &= \frac{1}{2^{n+2}}\sum\limits_{b\in\mathcal{O}\cap\{a,d\}^\perp}c_b[\Tr(T_b)+(-1)^s\Tr(T_aT_b)\nonumber\\
            &\hspace{25mm}+(-1)^t\Tr(T_dT_b)+(-1)^{s+t}\Tr(T_aT_dT_b)]\nonumber\\
        &= \frac{1}{4}\left[1+(-1)^sc_a+(-1)^tc_d+(-1)^{s+t+\beta(a,d)}c_{a+d}\right]\nonumber\\
        &= \frac{(-1)^t}{4}(c_d+(-1)^{s+\beta(a,d)}c_{a+d}).
    \end{align}
    Here in the first line we use idempotence of the projector $\Pi_d^t$ and the cyclic property of the trace, in the second line we use Eq.~\eqref{Equation:PauliUpdateExpansion}, in the third line we expand the projectors and simplify, in the fourth line we use the orthogonality relations of the Pauli operators, and in the last line we use the assumption that $c_a=(-1)^{s+1}$.
    
    That is, $\Tr(\Pi_d^1\Pi_a^sA_{\mathcal{O}}^{c}\Pi_a^s)=-\Tr(\Pi_d^0\Pi_a^sA_{\mathcal{O}}^{c}\Pi_a^s)$. In order to not contradict the assumption $A_{\mathcal{O}}^{c}\in\Lambda$, we must have $\Tr(\Pi_d^0\Pi_a^sA_{\mathcal{O}}^{c}\Pi_a^s)=\Tr(\Pi_d^1\Pi_a^sA_{\mathcal{O}}^{c}\Pi_a^s)=0$. Then
    \begin{equation}
        \Tr(T_d\Pi_a^sA_{\mathcal{O}}^{c}\Pi_a^s)=\Tr(\Pi_d^0\Pi_a^sA_{\mathcal{O}}^{c}\Pi_a^s)-\Tr(\Pi_d^1\Pi_a^sA_{\mathcal{O}}^{c}\Pi_a^s)=0.
    \end{equation}
    This holds for all $d\in\mathcal{O}\cap a^\perp$.

    A similar argument shows that $\Tr(T_d\Pi_a^sA_{\mathcal{O}}^{c}\Pi_a^s)=0$ for all $d\in a+(\mathcal{O}\cap a^\perp)$. Therefore, the support of $\Pi_a^sA_{\mathcal{O}}^{c}\Pi_a^s$ is empty and so $\Pi_a^sA_{\mathcal{O}}^{c}\Pi_a^s=0$.
\end{proof}

\begin{Corollary}\label{Corollary:PauliClosure}
    $\bar{\mathcal{V}}$ is closed under Pauli measurements.
\end{Corollary}

\begin{proof}[Proof of Corollary~\ref{Corollary:PauliClosure}]
    This corollary follows from a combination of Lemma~\ref{Lemma:PauliClosureLG} along with the techniques used in the proof of Ref.~\cite[Theorem~3]{OkayRaussendorf2021}. Ref.~\cite[Theorem~3]{OkayRaussendorf2021} reduces a Pauli measurement on an operator of the form $g(A\otimes \ket{\sigma}\bra{\sigma})g^\dagger$ to measurement only on $A$, along with Clifford corrections. The statement of this theorem is for the case where the operator $A$ is a vertex of $\Lambda$, but it applies to our case as well and the proof is identical. This idea is closely related to the model of Pauli-based quantum computation~\cite{BravyiSmolin2016,PeresGalvao2021}.
    
    Consider a phase space point $(\mathcal{O},c)\in\bar{\mathcal{V}}$. By definition, the corresponding phase space point operator has the structure
    \begin{equation}
        A_{\mathcal{O}}^{c}=g\left(A_{\mathcal{O}'}^{c'}\otimes\ket{\sigma}\bra{\sigma}\right)g^\dagger
    \end{equation}
    for some Clifford operator $g\in\Cl$ and stabilizer state $\ket{\sigma}$, where $A_{\mathcal{O}'}^{c'}$ is a line graph operator in $\Lambda$. Applying the projection $\Pi_a^s$,
    \begin{align}
        \Pi_a^sA_{\mathcal{O}}^{c}\Pi_a^s =& \Pi_a^sg\left(A_{\mathcal{O}'}^{c'}\otimes\ket{\sigma}\bra{\sigma}\right)g^\dagger\Pi_a^s\nonumber\\
        =& g\Pi_{a'}^{s'}\left(A_{\mathcal{O}'}^{c'}\otimes\ket{\sigma}\bra{\sigma}\right)\Pi_{a'}^{s'}g^\dagger.
    \end{align}
    Here in the second line we use the defining property of Clifford operations---that they map Pauli operators to Pauli operators under conjugation---to map $g$ past $\Pi_a^s$.

    The Pauli observable $T_{a'}$ can be decomposed as $T_{a'}\propto T_{a_1}\otimes T_{a_2}$ where $T_{a_1}$ acts only on the subspace of the Hilbert space supporting $A_{\mathcal{O}'}^{c'}$ and $T_{a_2}$ acts only on the subspace supporting $\ket{\sigma}$. Here we have two cases.

    Case 1: If $\pm T_{a_2}$ is in the stabilizer group of $\ket{\sigma}$, the part of $T_{a'}$ acting on the second subspace can be replaced by the eigenvalue of $T_{a_2}$. In more detail, suppose $(-1)^{\nu}T_{a_2}$ is in the stabilizer of the stabilizer state $\ket{\sigma}$. Then,
    \begin{align}
        \Pi_{a'}^{s'}&\left(A_{\mathcal{O}'}^{c'}\otimes\ket{\sigma}\bra{\sigma}\right)\Pi_{a'}^{s'}\nonumber\\
        =& A_{\mathcal{O}'}^{c'}\otimes\ket{\sigma}\bra{\sigma}+(-1)^{s'}\left[\left(T_{a_1}A_{\mathcal{O}'}^{c'}\otimes T_{a_2}\ket{\sigma}\bra{\sigma}\right)\right.\nonumber\\
        &\left.+\left(A_{\mathcal{O}'}^{\vec{c'}}T_{a_1}\otimes\ket{\sigma}\bra{\sigma}T_{a_2}\right)\right]+T_{a_1}A_{\mathcal{O}'}^{c'}T_{a_1}\otimes T_{a_2}\ket{\sigma}\bra{\sigma}T_{a_2}\nonumber\\
        =& A_{\mathcal{O}'}^{c'}\otimes\ket{\sigma}\bra{\sigma}+(-1)^{s'+\nu}\left[\left(T_{a_1}A_{\mathcal{O}'}^{c'}\otimes\ket{\sigma}\bra{\sigma}\right)\right.\nonumber\\
        &\left.+\left(A_{\mathcal{O}'}^{\vec{c'}}T_{a_1}\otimes\ket{\sigma}\bra{\sigma}\right)\right]+T_{a_1}A_{\mathcal{O}'}^{c'}T_{a_1}\otimes\ket{\sigma}\bra{\sigma}\nonumber\\
        =&\Pi_{a_1}^{s'+\nu}A_{\mathcal{O}'}^{c'}\Pi_{a_1}^{s+\nu}\otimes\ket{\sigma}\bra{\sigma}.
    \end{align}
    Now we can apply Lemma~\ref{Lemma:PauliClosureLG}.

    Case 2: If $\pm T_{a_2}$ is not in the stabilizer of $\ket{\sigma}$, then $T_{a_2}$ anticommutes with some Pauli operator in the stabilizer of $\ket{\sigma}$ and the outcome $s'$ is uniformly random. Consequently, there exists a Clifford element $g_2\in\Cl$ (depending on $a'$ and $\sigma$) such that $g_2 T_{a'} g_2^\dagger = Z_{n-k+1}$ and $g_2^\dagger \left( A_{\mathcal{O}'}^{c'} \otimes \ket{\sigma}\bra{\sigma} \right) g_2 = A_{\tilde{\mathcal{O}}'}^{\tilde{c}'} \otimes \Pi^0_{x} \otimes \ket{\sigma '}\bra{\sigma '}$. Then we have
    \begin{align}
        \Pi_{a'}^{s'}&\left(A_{\mathcal{O}'}^{c'}\otimes\ket{\sigma}\bra{\sigma}\right)\Pi_{a'}^{s'}\nonumber\\
        =&\:g_2^\dagger \Pi^{s'}_{z_{n-k+1}} \left(A_{\tilde{\mathcal{O}}'}^{\tilde{c}'} \otimes \Pi_x^0 \otimes \ket{\sigma '}\bra{\sigma '} \right) \Pi^{s'}_{z_{n-k+1}} g_2.
    \end{align}
    Up to normalization this last expression is equivalent to
    \begin{align}
        g_2^\dagger \left(HX^{s'}\right)_{n-k+1}\left(A_{\tilde{\mathcal{O}}'}^{\tilde{c}'} \otimes \Pi_x^0 \otimes \ket{\sigma '}\bra{\sigma '} \right)\left(X^{s'} H\right)_{n-k+1} g_2 \nonumber\\
        = g_2^\dagger \left(HX^{s'}\right)_{n-k+1}g_2\left(A_{\mathcal{O}'}^{c'}\otimes \ket{\sigma}\bra{\sigma} \right)g_2^\dagger\left(X^{s'} H\right)_{n-k+1} g_2
    \end{align}
    where $H$ is the Hadamard gate. This operator has the structure of Eq.~\eqref{Equation:phasespace}.
\end{proof}

\begin{Corollary}\label{Corollary:PauliClosureStates}
    The set of states positively representable by $\bar{\mathcal{V}}$ in Eq.~\eqref{Equation:LineDecomp} is closed under Pauli measurements.
\end{Corollary}

\begin{proof}[Proof of Corollary~\ref{Corollary:PauliClosureStates}]
    Suppose a state $\rho$ has a nonnegative decomposition in Eq.~\eqref{Equation:LineDecomp}. If a Pauli measurement $T_a$ is performed on $\rho$ giving measurement outcome $s\in\mathbb{Z}_2$ the post-measurement state is
    \begin{align}
        \frac{\Pi_a^s\rho\Pi_a^s}{\Tr(\Pi_a^s \rho)}=&\frac{1}{\Tr(\Pi_a^s \rho)}\sum_{(\mathcal{O},c)\in\mathcal{V}}W_\rho(\mathcal{O},c)\Pi_a^sA_{\mathcal{O}}^{c}\Pi_a^s\nonumber\\
        =&\sum_{\substack{(\mathcal{O}, c) \in \mathcal{V} \\ \Tr(\Pi_a^s A_{\mathcal{O}}^c) \neq 0}} \frac{\Tr(\Pi_a^s A_{\mathcal{O}}^c)}{\Tr(\Pi_a^s \rho)}W_{\rho}(\mathcal{O}, c) \cdot \frac{\Pi_a^s A_{\mathcal{O}}^c \Pi^s_a}{\Tr(\Pi_a^s A_{\mathcal{O}}^c)}.
    \end{align}
    Which gives a positive decomposition in terms of phase point operators in $\bar{\mathcal{V}}$ whenever $W_{\rho}(\mathcal{O},c)$ is positive.
\end{proof}

\subsection{Classical simulation algorithm}\label{Section:ClassicalSimulation}

As shown in Sections~\ref{Section:CliffordUpdateRules} and~\ref{Section:PauliUpdateRules}, the generalized phase space over which the quasiprobability representation is defined is closed under Clifford gates and Pauli measurements. This fact allows us to define a classical simulation algorithm for quantum computation with magic states that applies whenever the representation of the input state is nonnegative, Algorithm~\ref{Algorithm:ClassicalSimulation}. The proof of correctness for this algorithm is analogous to the proof of Theorem~5 of Ref.~\cite{RaussendorfZurel2020} and the proof of Theorem~2 of Ref.~\cite{ZurelRaussendorf2020}. With certain additional assumptions, this algorithm is also efficient, this is the result of the following theorem.

\begin{algorithm}[H]
	\begin{algorithmic}[1]
		\REQUIRE $p_{\rho}$, sequence of gates and measurements $\mathcal{C}$
        \ENSURE sequence of simulated measurement outcomes $\mathcal{M}$
		\STATE sample a point $(g, \ket{\sigma}, A_\mathcal{O}^c)\in\mathcal{V}$ according to the probability distribution $p_{\rho}$
		\WHILE{end of circuit has not been reached}
		      \IF{a Clifford gate $h\in\Cl$ is encountered}
		          \STATE update $(g, \ket{\sigma}, A_\mathcal{O}^c)\leftarrow (hg, \ket{\sigma}, A_\mathcal{O}^c)$
		      \ENDIF
		      \IF{a Pauli measurement $T_a ,\;a\in E$ is encountered}
                \STATE define $T_{a_1}\otimes T_{a_2} := T_{S_g a}$
                \IF{$T_{a_2} \ket{\sigma} = (-1)^\nu \ket{\sigma}$}
                    \STATE choose $s=0$ or $s=1$ with probability $\frac{1+(-1)^sc_{a_1}}{2}$
                    \STATE add $(-1)^{s + \nu + \Phi_g(a)}$ to $\mathcal{M}$
                    \STATE update $g \leftarrow g \cdot g_1$
                    \STATE update $\ket{\sigma} \leftarrow \ket{s} \otimes \ket{\sigma}$
                    \STATE update $A_O^c \leftarrow g_1^\dagger\cdot \left( A^{\mathcal{O}\cap a_1^\perp}_{c|_{\mathcal{O}\cap a_1^\perp}}\right) \cdot g_1$
                \ELSE
                    \STATE choose $s = 0$ or $s=1$ with equal probability
                    \STATE add $(-1)^{s + \Phi_g(a)}$ to $\mathcal{M}$
                    \STATE update $g \leftarrow g\cdot g_2^\dagger\cdot \left(HX^{s}\right)_{n-k-1} \cdot g_2$
                \ENDIF
		      \ENDIF
		\ENDWHILE
	\end{algorithmic}
    \caption{One run of the classical simulation of quantum computation with magic states based on the quasiprobability representation defined in Sections~\ref{Section:PhaseSpaceDescription} and the update rules described in Sections~\ref{Section:CliffordUpdateRules} and~\ref{Section:PauliUpdateRules}. We describe points in $\mathcal{V}$ (and $\bar{\mathcal{V}}$) by a triple in accordance with Eq.~\eqref{Equation:phasespace} and $g_1$ and $g_2$ are defined as in Lemma~\ref{Lemma:PauliClosureLG} and Corollary~\ref{Corollary:PauliClosure} respectively. The algorithm provides samples from the joint probability distribution of the Pauli measurements in a quantum circuit consisting of Clifford gates and Pauli measurements applied to an input state $\rho$ such that $W_\rho \geq 0$.}
    \label{Algorithm:ClassicalSimulation}
\end{algorithm}

\begin{Theorem}\label{Theorem:SimulationEfficiency}
    The classical simulation given in Algorithm~\ref{Algorithm:ClassicalSimulation} is efficient on all input states $\rho$ for which there exists a decomposition Eq.~\eqref{Equation:LineDecomp} with nonzero coefficients $W_\rho(\mathcal{O},c)$, and for which samples can be efficiently obtained from the distribution $W_\rho$.
\end{Theorem}

\begin{proof}[Proof of Theorem~\ref{Theorem:SimulationEfficiency}]
    In order to prove efficiency of the classical simulation algorithm, we must prove (I)~existence of an efficient representation of each component (i.e. phase space points, Clifford operations, and Pauli measurements), and (II)~efficiently computable updates of the phase space points under Clifford operations and Pauli measurements.

    (I)~Pauli observables are represented by $2n$ bit strings (possibly with an extra bit indicating a sign) as in Eq.~\eqref{Equation:PauliOperator}. Clifford operations can be specified by a $2n\times2n$ bit matrix indicating the action of the Clifford on a basis of the labels of the Pauli operators, $\mathbb{Z}_2^{2n}$, together with an extra $2n$ bits indicating the signs of the Pauli operators.\footnote{This description of a Clifford operation contains some redundant information as the constraint that the Clifford operation must preserve the commutation relations of the Pauli operators constrains the binary matrix to be a symplectic map on $E$, but it suffices for our purposes.} Algorithms for converting between unitary Pauli and Clifford operators and this representation exist~\cite{deSilvaYin2024}. In the simulation, we represent phase space points as in Eq.~\eqref{Equation:phasespace}, by a triple $(g,\ket{\sigma},A_{\mathcal{O}}^c)$, where $g\in\Cl$ is a Clifford group element, $\ket{\sigma}$ is a stabilizer state, and $A_{\mathcal{O}}^c$ is a line graph operator in $\Lambda$. As above, specifying the Clifford operator requires a $2n\times2n$ binary matrix together with an additional $2n$ bits. The stabilizer state $\ket{\sigma}$ can be specified by the generators of its stabilizer group, which requires $2n^2+n$ bits, namely, $2n^2$ bits to specify the $n$ Pauli operators that generate the group, and another $n$ bits for their signs. Finally, specifying $A_{\mathcal{O}}^c$ requires specifying its support $\mathcal{O}$ and the coefficients $c$. Every line graph operator has at most $O(n^2)$ nonzero coefficients when expanded in the Pauli basis~\cite{ChapmanPrivate,YucelPrivate}. Therefore, specifying $\mathcal{O}$ requires $O(n^3)$ bits. The coefficients $c$ are $O(n^2)$ real numbers, approximation of the coefficients by bits is a matter of numeric precision.
    
    (II)~Proving efficiency of the the updates of phase space points under Clifford gates and Pauli measurements relies on the structure of the phase space points in Eq.~\eqref{Equation:phasespace}. These operators have a stabilizer-like part consisting of the Clifford operation $g$ and the stabilizer state $\ket{\sigma}$, and a line graph operator $A_{\tilde{\mathcal{O}}}^{\tilde{c}}$. These two parts can be tracked separately in the simulation. Under Clifford gates, the line graph operator part is unaffected, and update of the stabilizer part follows from standard stabilizer techniques. Under Pauli measurements, using now standard stabilizer techniques~\cite{BravyiSmolin2016,OkayRaussendorf2021,PeresGalvao2021} (described in the previous section), the measurement can be efficiently reduced to a measurement only on the line graph part. Efficient update of this part follows from the fact that the line graph property restricts the number of terms that can appear when the operator is expanded in the Pauli basis, as alluded to above. Some additional details for some of the steps are provided below.
    
    We consider a circuit on an initial $n$-qubit state consisting of $m$ Clifford gates and $m'$ Pauli measurements. Whenever a Clifford gate is encountered, for the update in line~4 of Algorithm~\ref{Algorithm:ClassicalSimulation}, we must compute the composition of two $n$-qubit Clifford gates. This requires two steps: matrix multiplication in $\mathbb{Z}_2^{2n}$ and $O(n^2)$ evaluations of the $\beta$ function in Eq.~\eqref{Equation:BetaDefinition} to determine the updated signs of the Clifford operation. The time complexity of each of these steps is polynomial in $n$.

    If a Pauli measurement $a\in E$ is encountered we proceed in several steps outlined in lines~6--19 of Algorithm~\ref{Algorithm:ClassicalSimulation}. First, we compute $S_ga$, and find the subvectors $a_1$ and $a_2$. This involves matrix-vector multiplication in $\mathbb{Z}_2^{2n}$, which has time complexity $O(n^2)$, and a partitioning of the vector into two subvectors depending on the number of qubits supporting the stabilizer state $\ket{\sigma}$. Determining whether $\pm T_{a_2}$ is in the stabilizer group of $\ket{\sigma}$ is a matter of determining whether $a_2$ is in the span of the Pauli generators of the stabilizer group. This is again a linear algebra problem in $\mathbb{Z}_2^{2n}$ with that can be accomplished using Gaussian elimination (i.e. time complexity $O(n^3)$). In the case where $\pm T_{a_2}$ is in the group, $\nu$ can be determined by recursive use of the relation Eq.~\eqref{Equation:NoncontextualityCondition} on the signs of the generators of the stabilizer group.

    The remaining steps can be easily filled in. They involve linear algebra subroutines in $\mathbb{Z}_2^{2n}$, each with time complexity polynomial in $n$. The number of the linear algebra subroutines that need to be performed is polynomial in $m$, $m'$, and $n$. Therefore, assuming samples can be efficiently obtained from the distribution representing the input state of the circuit, the simulation algorithm is efficient.
\end{proof}

In general, a state $\rho$ will not admit a decomposition of the form Eq.~\eqref{Equation:LineDecomp} such that $W_\rho (\mathcal{O}, c) \geq 0$ for all $(\mathcal{O}, c)$. Consequently, we define the robustness of a state as
\begin{equation}
    \mathfrak{R}(\rho) := \min\limits_{W}\left\{||W||_1\;\bigg|\;\rho=\sum\limits_{(\mathcal{O},c)\in\mathcal{V}}W(\mathcal{O},c)A_{\mathcal{O}}^{c}\right\}.
\end{equation}
Since line graph operators are preserved under Clifford gates and Pauli measurement, the robustness is a monotone with respect to the resource theory of stabilizer quantum computation~\cite{VeitchEmerson2014}. The cost of classical simulation in the presence of negativity can then be related to this robustness~\cite{PashayanBartlett2015}. This generalized robustness is bounded above by the phase-space robustness of the CNC construction~\cite{RaussendorfZurel2020} and the robustness of magic~\cite{HowardCampbell2017}. See Appendix~\ref{Appendix:Robustness} for details.

\section{New vertices of the \texorpdfstring{$\Lambda$}{Lambda} polytopes}\label{Section:NewLambdaVertices}

The generalized phase space of the present model is defined in part by looking at intersections of the $\Lambda$ polytopes. Depending on the subspace with which we intersect, the resulting polytope may share vertices with $\Lambda$. In this section we show that this is indeed the case. In particular, we show that, for any number $n$ of qubits, intersecting $\Lambda$ with the space of operators with support $\mathcal{O}$ for which the frustration graph of $\mathcal{O}^*$ is $L(K_{2n+1})$---the line graph of the complete graph on $2n+1$ vertices---gives a polytope which shares vertices with $\Lambda$. This is the content of Theorem~\ref{Theorem:NewLineGraphVertices} below.

In this case we can determine the coefficients of the vertices in the Pauli basis as well, thus we obtain a complete characterization of new families of vertices of the $\Lambda$ polytopes for every number of qubits. By polar duality~\cite{Heimendahl2019,ZurelHeimendahl2021}, we also obtain a complete characterization of new families of facets of the stabilizer polytope for every number of qubits.

\begin{Theorem}\label{Theorem:NewLineGraphVertices}
	Define the operator
	\begin{equation}\label{Equation:NewLineGraphVertices}
		A_{\mathcal{O}}^{\eta} = \frac{1}{2^n} \left( 1 + \frac{1}{n} \sum_{b\in\mathcal{O}^*} (-1)^{\eta(b)} T_b \right)
	\end{equation}
	where $\mathcal{O}^*$ is any set of Pauli operators whose frustration graph is $L(K_{2n+1})$. Then there exist choices for the signs $\eta:\mathcal{O}^*\rightarrow\mathbb{Z}_2$ such that the operators $A_\mathcal{O}^\eta$ of the form Eq.~(\ref{Equation:NewLineGraphVertices}) are vertices of $\Lambda$.
\end{Theorem}
The rest of this section is devoted to the proof of this theorem.

\begin{proof}[Proof of Theorem~\ref{Theorem:NewLineGraphVertices}]
By Theorem~18.1 of Ref.~\cite{Chvatal1983}, to prove that an operator $A_{\mathcal{O}}^\eta$ is a vertex of $\Lambda$, it suffices to show (1) that $A_{\mathcal{O}}^\eta$ is in $\Lambda$, and (2) that the set of projectors onto stabilizer states which are orthogonal to $A_{\mathcal{O}}^\eta$ with respect to the Hilbert-Schmidt inner product has rank $4^n-1$ when viewed as vectors of Pauli basis coefficients.

To start, we can directly compute the inner product $\Tr(\Pi_I^rA_{\mathcal{O}}^\eta)$ for any choice of signs $\eta$ and any projector onto a stabilizer state $\Pi_I^r$:
\begin{align}
    \Tr(\Pi_I^rA_{\mathcal{O}}^\eta)=&\frac{1}{2^n}+\frac{1}{n2^{2n}}\sum\limits_{a\in I}\sum\limits_{b\in \mathcal{O}^*}(-1)^{r(a)+\eta(b)}\Tr(T_aT_b)\nonumber\\
    =&\frac{1}{2^n}+\frac{1}{n2^n}\sum\limits_{a\in I\cap\mathcal{O}^*}(-1)^{r(a)+\eta(a)}.
\end{align}
Since the largest independent set in $L(K_{2n+1})$ has size $n$, this inner product is always nonnegative and so $A_{\mathcal{O}}^{\eta}\in\Lambda$. Also, the inner product is zero if and only if $|I\cap\mathcal{O}^*|=n$ and $r(a)\ne\eta(a)$ for all $a\in I\cap\mathcal{O}^*$. All that remains is to show that the signs $\eta$ can be chosen so that the set of stabilizer states for which this inner product is zero has rank $4^n-1$.

\begin{figure}
    \centering
    \includegraphics[width=0.5\textwidth]{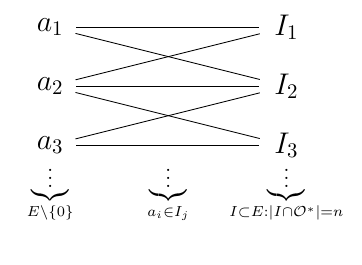}
    \caption{A sketch of a bipartite graph describing the inclusion relations of elements $a_i\in E\setminus\{0\}$ in a set of isotropic subspaces $I_j\subset E$ each with the property that $|I_j\cap\mathcal{O}^*|=n$. An edge connects an observable $a_i$ on the left to an isotropic subspace $I_j$ on the right if and only if $a_i\in I_j$.}
    \label{Figure:New vertices proof - Bipartite graph}
\end{figure}

Consider the bipartite graph $G$ constructed as follows: $G$ has a vertex for each Pauli observable $a\in E\setminus\{0\}$, a vertex for each isotropic subspace $I\subset E$ such that $|I\cap\mathcal{O}^*|=n$, and an edge connecting a vertex $a\in E$ to an isotropic subspace $I\subset E$ if and only if $a\in I$. A sketch of this graph is shown in Fig.~\ref{Figure:New vertices proof - Bipartite graph}. We need to show that there exists a choice of signs for the edges of this graph such that each off-diagonal block of the corresponding signed adjacency matrix has rank $4^n-1$.

Using a graph theoretic result proven in Appendix~\ref{Appendix:GraphTheoryResults} (see in particular Corollary~\ref{Corollary2}), to establish this it suffices to show that the graph $G$ has a matching of size $4^n-1$. By K\H{o}nig's theorem, this is true if and only if the minimum vertex cover of $G$ has order $4^n-1$. One vertex cover of order $4^n-1$ is obtained by taking all vertices on the Pauli operator side of the bipartition. Now we must show that a smaller vertex cover cannot be obtained by removing vertices from the Pauli side and adding fewer vertices on the isotropic subspace side. To do this we compute the degree of each vertex in the graph. If we can show that the degree of each vertex on the left hand side of the graph in Fig.~\ref{Figure:New vertices proof - Bipartite graph} is larger than the degree of every vertex on the right hand side of the graph, then the minimum vertex cover has size $4^n-1$ and the result is proved.

Each isotropic subspace $I\subset E$ has order $2^n$, one of these elements is $0$, and so the degree of each vertex on the right hand side is $2^n-1$. Now we need to count the number of isotropic subspaces $I\subset E$ containing each observable $a\in E\setminus\{0\}$. Here we have many cases. Up to an overall phase, each Pauli observable $T_a$ can be written as a product of some subset of the $2n+1$ pair-wise anticommuting observables $C_1,C_2,\dots,C_{2n+1}$ which generate $\mathcal{O}$ by taking pair-wise products. Since the product of all $2n+1$ of these Pauli operators is proportional to the identity, this representation is not unique. Each Pauli observable $a$ will have exactly two factorizations of the form $T_a\propto C_{\mu_1}C_{\mu_2}\cdots C_{\mu_k}$, one where $k$ is odd and one where $k$ is even. Therefore, without loss of generality we can restrict our attention to the factorizations where $k$ is even.

Suppose $T_{a}\propto C_{\mu_1}C_{\mu_2}\cdots C_{\mu_{2m}}$. In order for $I\subset E$ to contain $a$, $I$ must contain a set of $m$ generators, each of which is proportional to a product of a pair of the operators $C_{\mu_1},C_{\mu_2},\dots$. Once these $m$ generators are specified, the remaining $n-m$ generators of $I$ must be chosen from products of pairs of operators from the remaining operators $\{C_1,C_2,\dots\}\setminus\{C_{\mu_1},C_{\mu_2},...\}$. Therefore, the number of isotropic subspaces $I$ containing $a$ is
\begin{widetext}
\begin{align}
    f(n,m)=&\left[\frac{1}{m!}\prod\limits_{j=0}^{m-1}\begin{pmatrix}2m-2j\\2\end{pmatrix}\right]\cdot\frac{1}{(n-m)!}\prod\limits_{k=0}^{n-m-1}\begin{pmatrix}2n+1-2m-2k\\2\end{pmatrix}\nonumber\\
    =&\frac{1}{m!(n-m)!2^n}\left[\prod\limits_{j=1}^{2m}j\right]\cdot\left[\prod\limits_{j=1}^{2n+1-2m}j\right]\nonumber\\
    =&\frac{(2m)!(2n-2m)!}{m!(n-m)!2^n}\cdot(2n+1-2m).\label{Equation:vertex degree function}
\end{align}
\end{widetext}
For this function, we can establish the bound $f(n,m)\ge2^n-1$ for all $n,m$. This proof of this bound is given in Lemma~\ref{Lemma:LowerBoundOnF} in Appendix~\ref{Appendix:LowerBoundOnF}. Therefore, each vertex on the left hand side of the graph in Figure~\ref{Figure:New vertices proof - Bipartite graph} has higher degree than every vertex on the right hand side, and so the minimum vertex cover has size $4^n-1$. This completes the proof.
\end{proof}

\section{Discussion}\label{Section:Discussion}

In this work, we presented a new quasiprobability representation for quantum computation with magic states based on generalized Jordan-Wigner transformations. We demonstrated that this representation has efficiently computable update rules with respect to Clifford gates and Pauli measurements. Moreover, it extends previous representations including those based on quasiprobabilistic decompositions in projectors onto stabilizer states~\cite{HowardCampbell2017}, and the CNC construction~\cite{RaussendorfZurel2020}. Using this new construction, we can efficiently simulate magic state quantum circuits on a larger class of input states than was previously known, thus pushing back the boundary between efficiently classically simulable and potentially advantageous quantum circuits.

This model has a close connection to the probabilistic model of quantum computation based on the $\Lambda$ polytopes. Namely, for each number $n$ of qubits, it defines a new polytope contained inside the $\Lambda$ polytope with some shared vertices. These vertices include the previously known CNC vertices~\cite{RaussendorfZurel2020,ZurelRaussendorf2020,Heimendahl2019}, as well as some new infinite families of vertices, as shown by Theorem~\ref{Theorem:NewLineGraphVertices}.

\paragraph*{Outlook.} Our results provide several avenues for future research. One possible direction is to look for more families of vertices of the $\Lambda$ polytopes which also have the line graph structure like those described in Theorem~\ref{Theorem:NewLineGraphVertices}.

A more speculative idea is to look at the potential application of these results for magic state distillation protocols. One of our results is a characterization of a new family of vertices of the $\Lambda$ polytope. By polar duality~\cite{ZurelHeimendahl2021}, this also gives a full characterization of new families of facets of the stabilizer polytopes for every number of qubits. Facets of the stabilizer polytope have in the past been linked to magic state distillation~\cite{Reichardt2009}. The CNC type facets~\cite{RaussendorfZurel2020,Heimendahl2019}, as well as the new facets defined by Theorem~\ref{Theorem:NewLineGraphVertices} could be used in a similar way for any number of qubits.

\section*{Acknowledgements}
\noindent M.Z. and R.R. are funded by the National Science and Engineering Research Council of Canada (NSERC), in part through the Canada First Research Excellence Fund, Quantum Materials and Future Technologies Program, and the Alliance International program. This work is also supported by the Horizon Europe project FoQaCiA, a European-Canadian collaboration [NSERC funding reference number 569582-2021]. L.Z.C. acknowledges support from the Australian Research Council through project number DP220101771 and the Centre of Excellence in Engineered Quantum Systems (EQUS) project number CE170100009. We thank Stephen Bartlett, Adrian Chapman, and Atak Talay Yucel for fruitful discussions.

\bibliography{refs}

\appendix
\onecolumngrid

\section{Graph theoretic lemmas}\label{Appendix:GraphTheoryResults}
In the proof of Theorem~\ref{Theorem:NewLineGraphVertices}, we use some concepts from graph theory. A signed graph $G^\sigma$ is an undirected graph $G$ together with a sign function $\sigma:\text{Edge}(G)\rightarrow\mathbb{Z}_2$ that assigns to each edge $e$ of $G$ a sign $(-1)^{\sigma(e)}$. We say a graph has full rank if its adjacency matrix has full rank. The perrank of a graph $G$ is the order of the largest subgraph of $G$ which is a disjoint union of copies of $K_2$ and cycles. If $G$ has order $N$ we say $G$ has full perrank if $perrank(G)=N$.

We use the following Lemma from Ref.~\cite{AkbariNahvi2020}
\begin{Lemma}[Ref.~\cite{AkbariNahvi2020}, Theorem~2.1]
    Let $G$ be a graph. Then there exists a sign function $\sigma$ for $G$ such that $G^\sigma$ has full rank if and only if $G$ has full perrank.
\end{Lemma}
In a bipartite graph, all cycles are even, so edges can be removed from any cycle of a bipartite graph to give copies of $K_2$ which cover the same vertices. Therefore, for bipartite graphs, the existence of an order $N$ spanning subgraph composed of a disjoint union of edges and cycles is equivalent to the existence of a matching of size $N$. Thus we get the following corollaries:
\begin{Corollary}
    Let $G$ be a bipartite graph with adjacency matrix
    \begin{equation}\label{Equation:Bipartite adjacency matrix}
        A = \begin{bmatrix}0&B\\B^T&0\end{bmatrix}.
    \end{equation}
    where $B$ is a square matrix. Then there exists a choice of signs $\sigma$ for $G$ so that $B$ has full rank if and only if there exists a perfect matching of $G^\sigma$.
\end{Corollary}

\begin{Corollary}\label{Corollary2}
    Let $G$ be a bipartite graph with adjacency matrix of the form Eq.~\eqref{Equation:Bipartite adjacency matrix} where $B$ is a $M\times N$ matrix with $M\le N$. Then there exists a choice of signs $\sigma$ for $G$ so that $B$ has rank $2M$ if and only if there exists a matching of order $M$.
\end{Corollary}

\section{Lower bound on \texorpdfstring{$f(n,m)$}{f(n,m)}}\label{Appendix:LowerBoundOnF}
Here we prove the lower bound on the function $f(n,m)$ defined in Eq.~\eqref{Equation:vertex degree function} and used in the proof of Theorem~\ref{Theorem:NewLineGraphVertices}. We restate the bound in the following lemma.

\begin{Lemma}\label{Lemma:LowerBoundOnF}
    For the function $f(n,m)$ defined in Eq.~\eqref{Equation:vertex degree function}, we have the bound $f(n,m)\ge2^n-1$.
\end{Lemma}

\begin{proof}[Proof of Lemma~\ref{Lemma:LowerBoundOnF}]
    The bound is established if we can prove the bound $\log_2(2^nf(n,m))\ge2n$. For this we use the exact form of Stirling's formula
    \begin{equation}
        \sqrt{2\pi n}\left(\frac{n}{e}\right)^ne^{\frac{1}{12n+1}}<n!<\sqrt{2\pi n}\left(\frac{n}{e}\right)^ne^{\frac{1}{12n}}.
    \end{equation}
    Applying this to $\log_2(2^nf(n,m))$, we have
    \begin{align}
        \log_2(2^nf(n,m))=&\log_2((2m)!)+\log_2((2n-2m)!)-\log_2(m!)-\log_2((n-m)!)+\log_2(2n+1-2m)\nonumber\\
        \ge & \hspace{4mm} \frac{1}{2}\log_2(2m) + 2m\log_2(2m) -2m\log_2(e) + \frac{\log_2(e)}{24m+1}\nonumber\\
        & +\frac{1}{2}\log_2(2(n-m)) + 2(n-m)\log_2(2(n-m)) - 2(n-m)\log_2(e) + \frac{\log_2(e)}{24(n-m)+1}\nonumber\\
        & -\frac{1}{2}\log_2(m) - m\log_2(m) + m\log_2(e) - \frac{\log_2(e)}{12m}\nonumber\\
        & -\frac{1}{2}\log_2(n-m) - (n-m)\log_2(n-m) +(n-m)\log_2(e) - \frac{\log_2(e)}{12(n-m)} + \log_2(2n+1-2m)\nonumber\\
        = & 1+2n + m\log_2(m) + (n-m)\log_2(n-m) - n\log_2(e) + \log_2(2n+1-2m)\nonumber\\
        &+\log_2(e)\left[\frac{1}{24m+1}+\frac{1}{24(n-m)+1}-\frac{1}{12m}-\frac{1}{12(n-m)}\right]\nonumber\\
        \ge & 2n + m\log_2(m) + (n-m)\log_2(n-m) - n\log_2(e) + 1 - \frac{\log_2(e)}{6}\nonumber\\
        \ge & 2n + m\log_2(m) + (n-m)\log_2(n-m) - n\log_2(e)
    \end{align}
    Here in the second last step we simply throw away some positive terms and use the fact $-\frac{1}{12m}-\frac{1}{12(n-m)}\ge-\frac{1}{6}$. In the last step we use $1-\log_2(e)/6>0$. This expression is minimized at $m=n/2$, and so we have
    \begin{equation}
        \log_2(2^nf(n,m)) \ge 2n + n(\log_2(n/2) - \log_2(e)).
    \end{equation}
    For $n\ge6$, $\log_2(n/2)>\log_2(e)$, and so we can establish the bound $\log_2(2^nf(n,m))\ge2n$. For the remaining cases $n<6$, we can compute the function and show directly that $f(n,m)\ge2^n-1$. These remaining cases are shown in Table~\ref{Table:New vertex proof - counting function small n}. Therefore, $f(n,m)\ge2^n-1$ for all $n,m$.

    \begin{table}
        \centering
        \begin{tabular}{|c||c|c|c|c|c||c|}
            \hline
            $f(n,m)\;\;n\backslash m$ & 1 & 2 & 3 & 4 & 5 & $2^n-1$\\
            \hline\hline
            1 &   1 &     &     &     &     & 1 \\
            \hline
            2 &   3 &   3 &     &     &     & 3 \\
            \hline
            3 &  15 &   9 &  15 &     &     & 7 \\
            \hline
            4 & 105 &  45 &  45 & 105 &     & 15 \\
            \hline
            5 & 945 & 315 & 225 & 315 & 945 & 31 \\
            \hline
        \end{tabular}
        \caption{The function $f(n,m)$ of Eq.~(\ref{Equation:vertex degree function}) evaluated for all $m$ and for $1\le n\le5$. This table, together with the calculation below show that $f(n,m)\ge2^n-1$ for all $n$ and $m$.}
        \label{Table:New vertex proof - counting function small n}
    \end{table}
\end{proof}

\section{Comparison with previous simulation algorithms}\label{Appendix:Robustness}

In this appendix, we compare the simulation algorithm introduced in this work with two prior approaches: (i) simulation based on quasimixtures of stabilizer states~\cite{HowardCampbell2017}, and (ii) the CNC construction~\cite{RaussendorfZurel2020}. We compare the approaches in two ways, in terms of the set of states that admit a nonnegative representation (and hence efficient classical simulation), and the robustness, a measure of the amount of negativity in the quasiprobability distribution, for states that do not admit a nonnegative representation.

Let $\mathcal{S}$ denote the set of pure stabilizer states on $n$ qubits and let $\text{STAB}=\text{Conv}(\{\ket{\sigma}\bra{\sigma}\;|\;\ket{\sigma}\in\mathcal{S}\})$ be the stabilizer polytope. We denote by $\mathcal{V}_{\mathrm{CNC}}$ the set of CNC phase space point operators (Eq.~\eqref{Equation:CNCOperators} in the main text), and by $\mathcal{V}$ the generalized phase space used in the present work, with the corresponding state representations
\begin{equation}\label{Equation:QuasiprobabilityRepresentations}
    \rho = \sum_{\ket{\sigma}\in\mathcal{S}}W_\rho(\ket{\sigma})\ket{\sigma}\bra{\sigma},
    \quad
    \rho = \sum_{\alpha\in\mathcal{V}_{\mathrm{CNC}}} W_{\rho}(\alpha)\,A_{\alpha},
    \quad\text{and}\quad
    \rho = \sum_{\beta\in\mathcal{V}} W_{\rho}(\beta)\,A_{\beta},
\end{equation}
respectively. For each classical simulation algorithm, the sets of states with nonnegative representations are
\begin{align}
    \mathcal{P}_{\mathrm{STAB}} &:= \text{STAB},\\
    \mathcal{P}_{\mathrm{CNC}} &:= \{\rho\;|\; \exists\, W\ge 0 \text{ in~\eqref{Equation:QuasiprobabilityRepresentations} over }\mathcal{V}_{\mathrm{CNC}}\},\\
    \mathcal{P}_{LG} &:= \{\rho\;|\; \exists\, W\ge 0 \text{ in~\eqref{Equation:QuasiprobabilityRepresentations} over }\mathcal{V}\}.
\end{align}
For states with no nonnegative representation, we quantify the cost of classical simulation using the robustness~\cite{PashayanBartlett2015,HowardCampbell2017},
\begin{align}
    \mathfrak{R}_{\mathrm{STAB}}(\rho) &:= \min\left\{\lVert W\rVert_1 \;\bigg|\; \rho=\sum_{\ket{\sigma}\in\mathcal{S}} W(\sigma)\ket{\sigma}\bra{\sigma}\right\},\\
    \mathfrak{R}_{\mathrm{CNC}}(\rho) &:= \min\left\{\lVert W\rVert_1 \;\bigg|\; \rho=\sum_{\alpha\in\mathcal{V}_{\mathrm{CNC}}} W(\alpha)A_{\alpha}\right\},\\
    \mathfrak{R}_{\mathrm{LG}}(\rho) &:= \min\left\{\lVert W\rVert_1 \;\bigg|\; \rho=\sum_{\beta\in\mathcal{V}} W(\beta)A_{\beta}\right\},
\end{align}
so that $\mathfrak{R}_{\mathrm{STAB}}$ is the usual robustness of magic~\cite{HowardCampbell2017} and $\mathfrak{R}_{\mathrm{LG}}$ is the generalized robustness used in this paper.

\subsection{States with nonnegative representations}

\begin{Lemma}\label{Lemma:positivity-sets}
    For $n\ge2$, the three positive regions satisfy the strict inclusions
    \begin{equation}\label{eq:positivity-inclusions}
        \mathcal{P}_{\mathrm{STAB}}\;\subsetneq\;\mathcal{P}_{\mathrm{CNC}}\;\subsetneq\;\mathcal{P}. 
    \end{equation}
\end{Lemma}

\begin{proof}[Proof of Lemma~\ref{Lemma:positivity-sets}]
(1) $\mathcal{P}_{\mathrm{STAB}}\subset\mathcal{P}_{\mathrm{CNC}}$: This inclusion follows from the fact that projectors onto stabilizer states are special cases of CNC phase space point operators. They correspond to phase space points $(\Omega,\gamma)$ with $\Omega=I$ a maximal isotropic subspace of $\mathbb{Z}_2^{2n}$. Hence any convex mixture of stabilizer states is CNC-nonnegative.

(2) $\mathcal{P}_{\mathrm{CNC}}\subset\mathcal{P}$: According to Eq.~\eqref{Equation:CNCOperatorAlt}, every CNC phase space point operator can be written as $A_{\Omega}^{\gamma}=g\left(A_{\tilde{\Omega}}^{\tilde{\gamma}}\otimes\ket{\sigma}\bra{\sigma}\right)g^{\dagger}$ with $\tilde{\Omega}^*$ pairwise anticommuting. The frustration graph of $\widetilde{\Omega}^{\ast}$ is $K_m$, which is the line graph of the star graph $K_{1,m}$. Hence $A_{\tilde{\Omega}}^{\tilde{\gamma}}$ is a line graph operator of the form~\eqref{Equation:phasespace}, and so $\mathcal{V}_{\mathrm{CNC}}\subset\mathcal{V}$. Therefore, any nonnegative CNC decomposition is also a nonnegative decomposition over $\mathcal{V}$.

(3) Strictness: Lemma~4 of Ref.~\cite{RaussendorfZurel2020} shows that $\mathcal{P}_{\mathrm{STAB}}\subsetneq\mathcal{P}_{\mathrm{CNC}}$. The frustration graph of the set of all Pauli operators on two qubits is a line graph. Therefore, every $2$-qubit state is in $\mathcal{P}_{\mathrm{LG}}$. Table~III of Ref.~\cite{RaussendorfZurel2020} shows that there are $2$-qubit states that are not in $\mathcal{P}_{\mathrm{CNC}}$. Let $\ket{\psi}$ be a $2$-qubit state in $\mathcal{P}_{\mathrm{LG}}\setminus\mathcal{P}_{\mathrm{CNC}}$, and let $\ket{\sigma}$ be any $n-2$-qubit stabilizer state. Then $\ket{\psi}\otimes\ket{\sigma}$ is an $n$-qubit state in $\mathcal{P}_{\mathrm{LG}}\setminus\mathcal{P}_{\mathrm{CNC}}$. Hence, for $n\ge2$ both inclusions are strict.
\end{proof}

\subsection{Robustness comparison}

\begin{Theorem}\label{Theorem:robustness-dominance}
    For all $n\ge1$ and all states $\rho$,
    \begin{equation}\label{Equation:robustness-chain}
        1 \;\le\; \mathfrak{R}_{\mathrm{LG}}(\rho) \;\le\; \mathfrak{R}_{\mathrm{CNC}}(\rho) \;\le\; \mathfrak{R}_{\mathrm{STAB}}(\rho).
    \end{equation}
\end{Theorem}

\begin{proof}[Proof of Theorem~\ref{Theorem:robustness-dominance}]
    The optimizations defining $\mathfrak{R}_{\mathrm{LG}}$, $\mathfrak{R}_{\mathrm{CNC}}$, and $\mathfrak{R}_{\mathrm{STAB}}$ are linear programs that minimize the $1$-norm over affine slices of the conic hulls of $\mathcal{V}$, $\mathcal{V}_{\mathrm{CNC}}$, and $\text{STAB}$, respectively. Because $\mathcal{S}\subset\mathcal{V}_{\mathrm{CNC}}\subset\mathcal{V}$ (see proof of Lemma~\ref{Lemma:positivity-sets}), feasible sets enlarge from right to left and the optimal values cannot increase, giving Eq.~\eqref{Equation:robustness-chain}.
\end{proof}

\subsection{Quantifying the gain using new \texorpdfstring{$\Lambda$}{Lambda}-vertices}

In this section we provide an upper bound on the robustness of magic~\cite{HowardCampbell2017} of the vertices described by Theorem~\ref{Theorem:NewLineGraphVertices}.

\begin{Lemma}\label{Lemma:RobustnessLowerBound}
    For every $n\ge 1$ and every vertex $A^{\eta}_{O}$ described in Theorem~3, the robustness of magic satisfies
    \begin{equation}
        \mathfrak{R}_S\left(A_{\mathcal{O}}^{\eta}\right) \le \frac{5n+2}{n}.
    \end{equation}
\end{Lemma}

\begin{proof}[Proof of Lemma~\ref{Lemma:RobustnessLowerBound}]
    We work in the Cartesian embedding of $\Lambda$ where any $X\in\mathrm{Herm}_1(\mathcal H)$ is represented by its vector of Pauli basis coefficients $x$ where
    \begin{equation}
        X=\frac{1}{2^n}\left(1+\sum\limits_{b\in\mathbb{Z}_2^{2n}\setminus\{0\}}x_bT_b\right).
    \end{equation}
    In these coordinates, the maximally mixed state corresponds to $x=0$, and stabilizer polytope facets have the form $1+v\cdot x\ge0$ for suitable normals $v$.

    Let $c$ be the coordinate vector of $A^{\eta}_{O}$, and let
    \begin{equation}
        H := \{x \;|\; 1 + c\cdot x = 0\}
    \end{equation}
    be the facet of the stabilizer polytope whose outward normal is $c$. Denote by $a$ the projection of $c$ onto $H$, and set $b:=0$. By construction, $a$, $b$, and $c$ are collinear and
    \begin{equation}\label{eq:a-proj}
        a=-\frac{c}{\lVert c\rVert_2^2}.
    \end{equation}
    Hence, there is an affine decomposition
    \begin{equation}\label{eq:two-point-decomp}
        c=ta + (1-t)b\quad\text{with}\quad t=-\lVert c\rVert_2^2.
    \end{equation}
    This corresponds to a decomposition of $A_{\mathcal{O}}^\eta$ as an affine combination of an operator of the form $\frac{1}{2^n}-\frac{1}{\lVert c\rVert_2^2}(A_{\mathcal{O}}^\eta-\frac{1}{2^n})$ that lies on the boundary of the stabilizer polytope, and the maximally mixed state. Since both of these operators are in the stabilizer polytope, they have nonnegative decomposition in projectors onto stabilizer states, and so this gives a decomposition of $A_{\mathcal{O}}^\eta$ as an affine combination of stabilizer states with coefficients that have $1$-norm $|t|+|1-t|$. This gives an upper bound on the robustness of magic of
    \begin{equation}
        \mathfrak{R}_{\mathrm{STAB}}\bigl(A^{\eta}_{O}\bigr) \le |t| + |1-t| \le 2\,\lVert c\rVert_2^2 + 1.
        \label{eq:l1-from-t}
    \end{equation}
    It remains to evaluate $\lVert c\rVert_2^2$ for the vertices of Theorem~3. We have
    \begin{equation}
        \Tr({A_{\mathcal{O}}^\eta}^2)=\frac{1}{2^n}+\frac{1}{2^n}\lVert c\rVert_2^2=\frac{1}{2^n}+\frac{1}{2^n}\cdot\frac{|\mathcal{O}|}{n^2},
    \end{equation}
    and so
    \begin{equation}
        \lVert c\rVert_2^2 = \frac{|\mathcal{O}|}{n^2} = \frac{2n+1}{n}.
        \label{eq:norm-c}
    \end{equation}
    Substituting \eqref{eq:norm-c} into \eqref{eq:l1-from-t} gives
    \begin{equation}
        \mathfrak{R}_{\mathrm{STAB}}\bigl(A^{\eta}_{O}\bigr)\;\le\; 2\cdot \frac{2n+1}{n} + 1\;=\; \frac{5n+2}{n},
    \end{equation}
    which is exactly the bound to be proven.
\end{proof}

\medskip

Our algorithm is a strict generalization of the CNC construction in the positivity sense (Lemma~\ref{Lemma:positivity-sets}), and its robustness never exceeds (and frequently improves upon) the CNC robustness (Theorem~\ref{Theorem:robustness-dominance}). For the new $\Lambda$ polytope vertices described by Theorem~\ref{Theorem:NewLineGraphVertices}, the robustness of magic obeys the upper bound in Lemma~\ref{Lemma:RobustnessLowerBound}, thus giving a modest improvement over stabilizer-based simulation.

\end{document}